\documentclass[12pt,a4paper]{article}
\usepackage{graphicx,fancyhdr}
\usepackage{cite,citesort}
\begin{document}

\pagenumbering{arabic}

\vspace{-40mm}

\begin{center}
\Large\textbf{Transient electrical conductivity of W-based
electron beam induced deposits during growth, irradiation and
exposure to air}\normalsize \vspace{8mm}

\large F. Porrati, R. Sachser and M. Huth

\vspace{5mm}

\small\textit{Physikalisches Institut, Goethe-Universit\"at,
Max-von-Laue-Str.~1, D-60438 Frankfurt am Main, Germany}

\end{center}

\vspace{0cm}
\begin{center}
\large\textbf{Abstract}\normalsize
\end{center}

W-based granular metals have been prepared by electron beam
induced deposition from the tungsten-hexacarbonyl W(CO)$_{6}$
precursor. \textit{In situ} electrical conductivity measurements
have been performed to monitor the growth process and to
investigate the behavior of the deposit under electron beam post
irradiation and by exposure to air. During the first part of the
growth process, the electrical conductivity grows non-linearly,
independent of the electron beam parameters. This behavior is
interpreted as the result of the increase of the W-particles
diameter. Once the growth process is terminated, the electrical
conductivity decreases with the logarithm of time, $\sigma \sim
ln(t)$. Temperature-dependent conductivity measurements of the
deposits reveal that the electrical transport takes place by means
of electron tunneling either between W-metal grains or between
grains and trap sites in the matrix. After venting the electron
microscope the electrical conductivity of the deposits shows a
degradation behavior, which depends on the composition. Electron
post-irradiation increases the electrical conductivity of the
deposits.

\begin{center}
\large\textbf{1. Introduction}\normalsize
\end{center}

Electron beam induced deposition (EBID) is a high resolution
one-step technique used to deposit and to pattern two- and
three-dimensional micro- and nano-structures\cite{koops}. The
importance of EBID is rapidly increasing in applied science and
fundamental research\cite{utke}. On the one hand, the possibility
of direct writing makes EBID a promising alternative to
nano-lithography and a useful tool for mask repair. On the other
hand, the capability to produce deposits from many different
precursors with tunable electrical properties makes this technique
attractive for the development of new materials. EBID is based on
the interaction of an electron beam with a substrate which is
covered by adsorbed precursor molecules and which contain the
metal or semiconductor to be deposited. The electrons dissociate
the precursor molecules into a volatile component, which leaves
the surface and into a non-volatile one, which forms the deposit.
The deposits consist of a disordered array of crystalline metallic
nanoparticles with diameters between about 1~nm to 5~nm embedded
in an insulating matrix. The metal volume fraction, i.e. the
average particle size and the interparticle distance, can be
varied by tuning the electron beam parameters (beam current,
acceleration voltage, dwell time).

$\textit{In situ}$ electrical conductivity measurements of
granular materials prepared by EBID or IBID (ion beam induced
deposition) are valuable in order to study the growth process and
the response of the material during post-irradiation and exposure
to air. By means of these measurements information about the
electrical transport properties, the microstructure, the chemical
and physical stability of the deposit are deduced. In literature,
$\textit{in situ}$ electrical measurements of IBID and EBID
deposits are rare\cite{utke}. The electrical behavior of EBID
deposits from acrylic acid has been monitored by means of
two-probe measurements\cite{utke}. Recently an ageing process has
been monitored in platinum-based nanostructures\cite{botman}. The
authors report a continuous decrease of the conductivity over a
time range of many days. Studies performed by using W(CO)$_{6}$ as
a precursor are known for deposits produced by
IBID\cite{prestigiacomo} and EBID\cite{hoyle,hoyle2}. The
investigation of Hoyle $\textit{ et al.}$\cite{hoyle,hoyle2}
represents an important reference for the present study. In their
work the authors prepared structures for beam energies between
2~keV and 20~keV. They investigated in situ the electrical
conductivity of the deposits and, by means of transmission
electron microscopy (TEM), their microstructure. It is the purpose
of the present paper to further investigate the properties of
W-based deposits from the W(CO)$_{6}$ precursor by performing
$\textit{in situ transient}$ electrical conductivity measurements
during deposition, post-irradiation and exposure to air.

\begin{center}
\large\textbf{2. Experimental}\normalsize
\end{center}

\begin{figure}\center{\includegraphics[width=8cm]{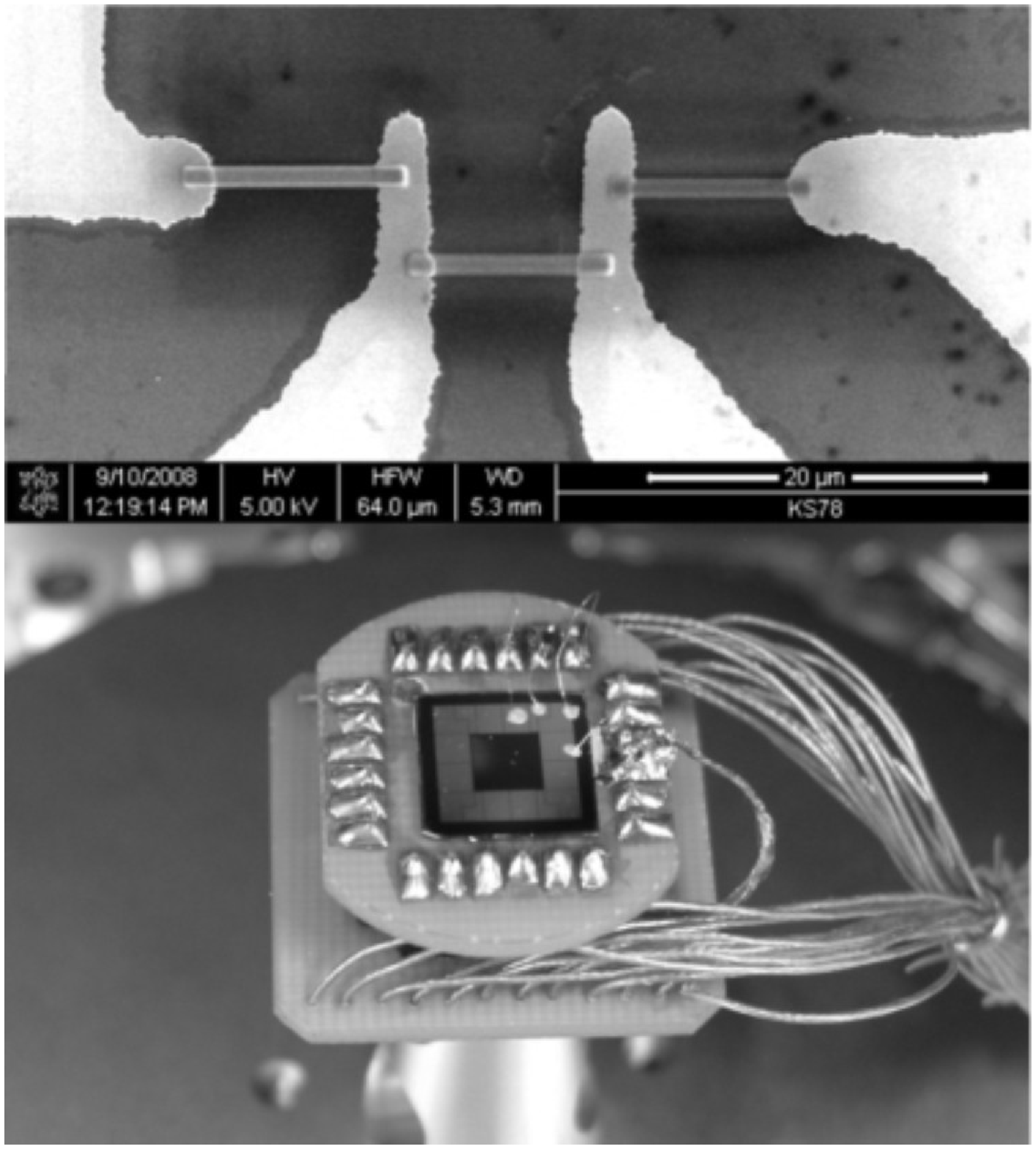}}
\caption{Upper figure: SEM image of three EBID structures for
two-probe electrical measurements. The deposits are written on
120~nm thick gold/chromium electrodes. Lower figure: Setup for the
$\textit{in situ}$ transient electrical conductivity measurements.
On this chip 3 from 12 possible measuring positions are contacted
via this copper wires attached to the contact pads with silver
paint.} \label{sem}\end{figure}

To prepare our samples we used a dual beam SEM/FIB microscope
(FEI, Nova Nanolab 600) with Schottky electron emitter and an
ultimate resolution of 1~nm. In this system the electron beam
power can be continuously tuned by means of the continuous
variation of the beam energy and pre-defined discrete values of
the beam current. The microscope is equipped with a gas injection
module which introduces the W(CO)$_{6}$ gas precursor via a 0.5~mm
diameter capillary in close proximity to the focus of the electron
or ion beam on the substrate surface. EBID structures were grown
on a Si (p-doped)/SiO$_{2}$ (300~nm) substrate. The substrates
were pre-patterned with 120~nm thick Au/Cr contacts defined by
UV-photolithography. In Fig.~\ref{sem} we show a scanning electron
microscope (SEM) image of three deposits for two-probe electrical
measurements. This technique was chosen after having verified that
the influence of the contact resistance between electrodes and
deposits is below about 3$\%$. Our chips are prepared to allow
measurements for up to 12 deposits. After deposition in situ
energy dispersive x-rays analysis (EDX) at 5~keV electron beam
energy was performed in order to determine the material
composition of the deposit. The low beam energy was chosen to
avoid excitation of x-ray fluorescence in the substrate material.
This was verified by Monte Carlo simulations of the electron
trajectories for the given thicknesses and compositions of the
deposit\cite{monte}. For in situ transient electrical conductivity
measurements a Keithley Sourcemeter coupled with a Multiplexer was
used to perform current measurements at fixed bias voltage. The
conductivity was deduced from the known dimensions of the
deposits. The lengths $\textit{l}$, the widths $\textit{w}$ and
the hight $\textit{h}$ of the samples were determined by direct
analysis of the SEM images. The maximum geometry-dependent error
for the conductivity data amounts to about 35$\%$. Finally,
temperature-dependent measurements of the electrical conductivity
where performed in a variable-temperature insert mounted in a
${^4}$He cryostat in the temperature range 1.8-265~K.

\begin{center}
\large\textbf{3. Measurements}\normalsize
\end{center}

\large\textbf{3.1 Deposition parameters and EDX characterization}\normalsize

\begin{figure}\center{\includegraphics[width=16cm]{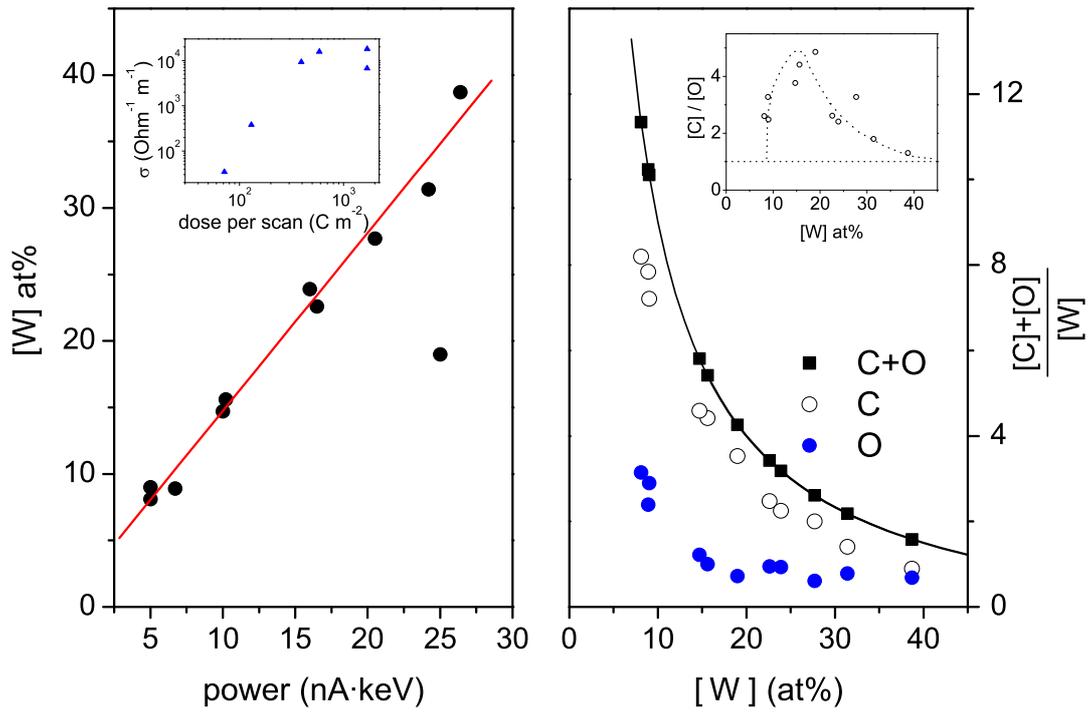}}
\caption{Left figure: Dependence of the W-metal content on the
electron beam power. The conductivity increases with the dose used
during deposition, see inset. Right figure: EDX analysis of
the composition of the samples prepared in this work. The
concentration of carbon and oxygen decrease with increasing
power of the electron beam. In the inset we report the ratio
between carbon and oxygen.} \label{composition}\end{figure}

In Fig.~\ref{composition} we report the results of the composition
analysis performed by means of EDX for the samples used in this
work. The deposits have been obtained with a variable voltage and
beam current in the range 4~keV$\leq$~E~$\leq$20~keV and
0.25~nA$\leq$~I~$\leq$6.6~nA, respectively. The corresponding beam
power varied between 5~nA$\cdot$keV$\leq$~p~$\leq$26~nA$\cdot$keV.
Within this range the deposits are granular with a W-content which
increases linearly between 8.1~at$\%$ and 38.7~at$\%$. In the left
inset of Fig.~\ref{composition} we plot the conductivity vs.~the
dose per scan. For high doses per scan the conductivity lies in
the range between ca. 9300 to 18000~$\Omega^{-1}$m$^{-1}$. These
values are in agreement with the ones reported by Hoyle
$\textit{et al.}$\cite{hoyle,hoyle2}. For smaller beam power we
measure a strong decrease of the conductivity which reaches
17~$\Omega^{-1}$m$^{-1}$ at 65~C/m${^2}$. This value is one order
of magnitude smaller than the one reported in
Ref.~\cite{hoyle,hoyle2} for comparable dose per scan. On the
right side of Fig.~\ref{composition} we plot the ratio between
oxygen (or carbon) and tungsten ([O]/[W] or [C]/[W]) content vs.
the tungsten-content for each deposit. The ratio ([O]+[C])/[W] is
also plotted. For W content equal to 8.1~at$\%$ we obtain
([O]+[C])/[C]=11.4. This value is close to the one deducible from
the stoichiometry of W(CO)$_{6}$ if one were to assume that the CO
molecules do not dissociate. By increasing the beam power the W
content increases, whereas the carbon and oxygen content decrease.
Additional information can be deduced from the ratio [C]/[O] (see
inset). For low doses we find [C]/[O]$>$1, which shows that the
composition of the matrix is dominated by carbon. The [C]/[O]
ratio shows a peak for a W content of ca.~15~at$\%$ and then it
decreases towards 1. By means of TEM measurements Hoyle
$\textit{et al.}$\cite{hoyle} found an amorphous material for
deposits prepared with a dose per scan smaller than 500~C/m$^{2}$.
For higher doses they found W containing nanocrystals consistent
with the high temperature $\beta$-phase of tungsten carbide
($\beta$-WC$_{1-x}$). The nanocrystal size was estimated to be
less than about 3~nm. From the EDX analysis of our deposits we
point out that the [W] content increases with increasing beam
power or with increasing dose per scan.

\large\textbf{3.2 Measurements during growth}\normalsize

\begin{figure}\center{\includegraphics[width=16cm]{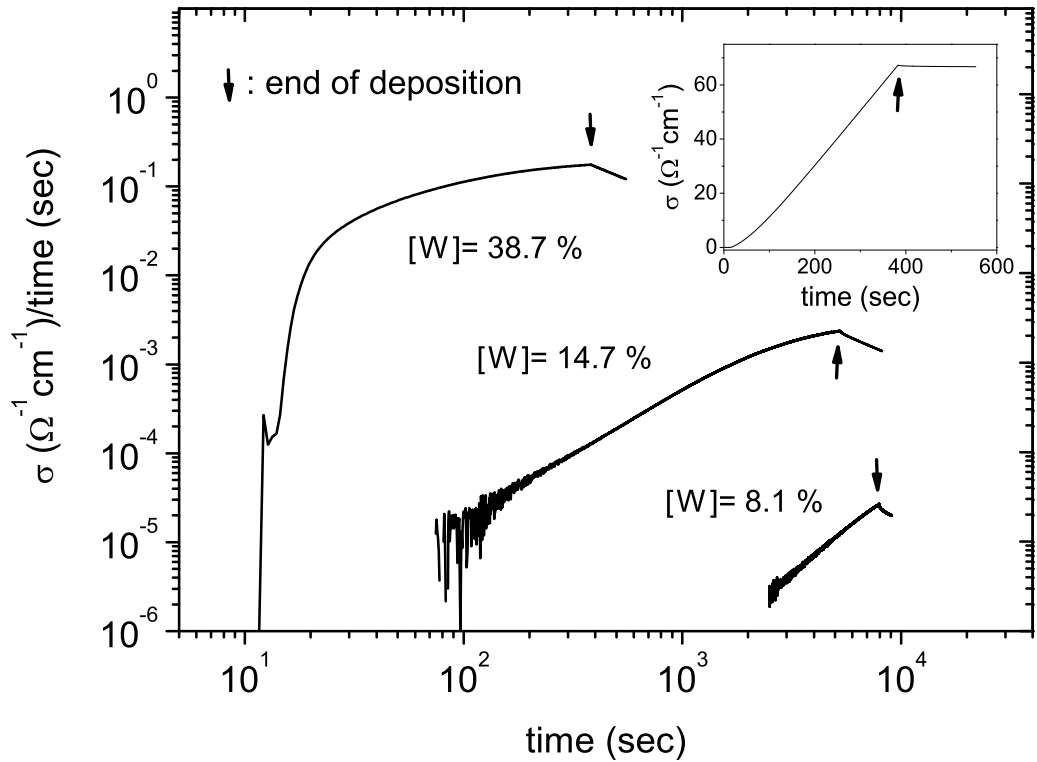}}
\caption{Electrical conductivity measured during the growth of
sample $\#$1 (8.1~$at\%$ W), $\#$4 (14.7~$at\%$) and $\#$9
(38.7~$at\%$), respectively. In each case, the deposition was
started at t=10~sec. The data points in the inset for sample $\#$9
show that after an initial non-linear growth, the conductivity
tends to increase linearly with the time. The conductivity of
samples $\#$1 and $\#$4 is non-linear during the whole deposition
time considered, which is due to the lower growth rate of these
deposits. The electrical conductivity of each sample decreases
once the deposition process is terminated.}
\label{growth_log}\end{figure}

In Fig.~\ref{growth_log} we show transient electrical conductivity
measurements during the growth of deposits with a W content of
8.1~$at\%$ (sample $\#$1), 14.7~$at\%$ (sample $\#$4) and
38.7~$at\%$ (sample $\#$9). For details concerning the
composition, the geometry and the beam parameters used during
deposition see Tab.~\ref{tab}.

\begin{table}[ht!]
\caption{Composition and deposition parameters of the samples.
Sample geometry (l/$\mu$m~$\times$~w/$\mu$m~$\times$~h/nm), beam
parameters (dwell~time/$\mu$s, pitch/nm): Sample~$\#1$
($8.75\times1.03\times240$), (100, 20); $\#2$
($9.19\times1.03\times315$), (100, 20); $\#3$
($8.27\times1.06\times226$), (100, 20); $\#4$
($4.76\times1.09\times228$), (100, 20); $\#5$
($6.15\times1.03\times148$), (100, 20); $\#6$
($18.1\times1\times352$), (100, 20); $\#7$
($4.57\times1.03\times241$), (100, 20); $\#8$
($14.93\times1.63\times192$), (100, 20); $\#9$
($7.75\times1.51\times682$), (5, 20).} \vspace{5mm} \centering
\begin{tabular}{c c c c c c}
\hline\hline
  Sample & W [$at\%$] & C [$at\%$] & O [$at\%$] & Energy/keV & Current/nA  \\
  \hline
  $\#$1 &  8.1 & 66.4 & 25.5 & 20 & 0.25 \\
  $\#$2 &  8.9 & 69.8 & 21.3 & 24 & 0.28 \\
  $\#$3 &  9.0 & 64.9 & 26.1 & 20 & 0.25 \\
  $\#$4 & 14.7 & 67.5 & 17.9 & 20 & 0.5 \\
  $\#$5 & 15.6 & 68.9 & 15.6 & 20 & 0.51 \\
  $\#$6 & 16.7 & 66.7 & 16.6 & 11 & 2.3 \\
  $\#$7 & 31.5 & 21.1 & 47.4 & 17 & 1.52 \\
  $\#$8 & 31.8 & 44.4 & 23.8 & 5 & 3.7 \\
  $\#$9 & 38.7 & 34.7 & 26.6 & 4 & 6.6 \\

  \hline
\end{tabular}
\label{tab}
\end{table}

At the time $t=10$~sec the gas precursor enters the vacuum
chamber. The e-beam is rastered over a rectangular area defining
the later deposit. In order to prepare samples $\#$1 and $\#$4 we
used a dwell time of 100~$\mu$s per pixel and a pitch of 20~nm.
For sample $\#$9 we used a dwell time of 5~$\mu$s per pixel and a
pitch of 20~nm. The conductivity changes at a rate $\sigma$'=
$\sigma$/t which increases monotonically with time during the
deposition of all samples. Sample $\#$9 shows a rapid increase of
$\sigma$' in the first seconds of deposition, which we attribute
to the formation of the first layer of the deposit. In the first 5
seconds $\sigma$' follows a power law of the form
$\sigma$'~=~$\textit{t}$~${^\alpha}$, with $\alpha$~=~3. After
that, $\sigma$' drastically decreases tending to become constant,
as can be expected, if the thickness of the deposit grows linearly
with time. $\sigma$' for the deposits with 8.1~$at\%$ and
14.7~$at\%$ metal content follows a power law of the form
$\sigma$'~$\sim$~$\textit{t}$~${^\alpha}$, with $\alpha$~=~1.86
and $\alpha$~=~1.43, respectively. For sample $\#$1 $\sigma$'
follows this power law during the whole deposition process, for
the sample $\#$4 it shows a decrease after about t~=~10$^{3}$~s.
The fit to sample $\#$4 was made between 300 and 1000 seconds. In
this range the growth rate is lower than in the first 300 seconds
of deposition, which we excluded from the fit because of the high
dispersion of the experimental points. Therefore it is not
surprising that the result of the fit gives for $\alpha$ a value
larger for sample $\#$1 than for sample $\#$4, which is
counterintuitive since we expect higher growth rate for deposits
with higher metal content.

In conclusion, these results clearly indicate that a linear
increase of the thickness is not sufficient to explain the
transient electrical conductivity measurements during the growth
of the deposits. Since the increase of the conductivity is faster
than linear, an additional mechanism has to be considered in order
to understand this observation. Among the possible mechanisms
which may contribute to the enhancement of $\sigma$' we consider
the charging effect and the increase of the metal grain size due
to the electron beam irradiation. In the last chapter of the paper
we discuss the relevance of these mechanisms for the samples
prepared in this work.

\begin{figure}\center{\includegraphics[width=16cm]{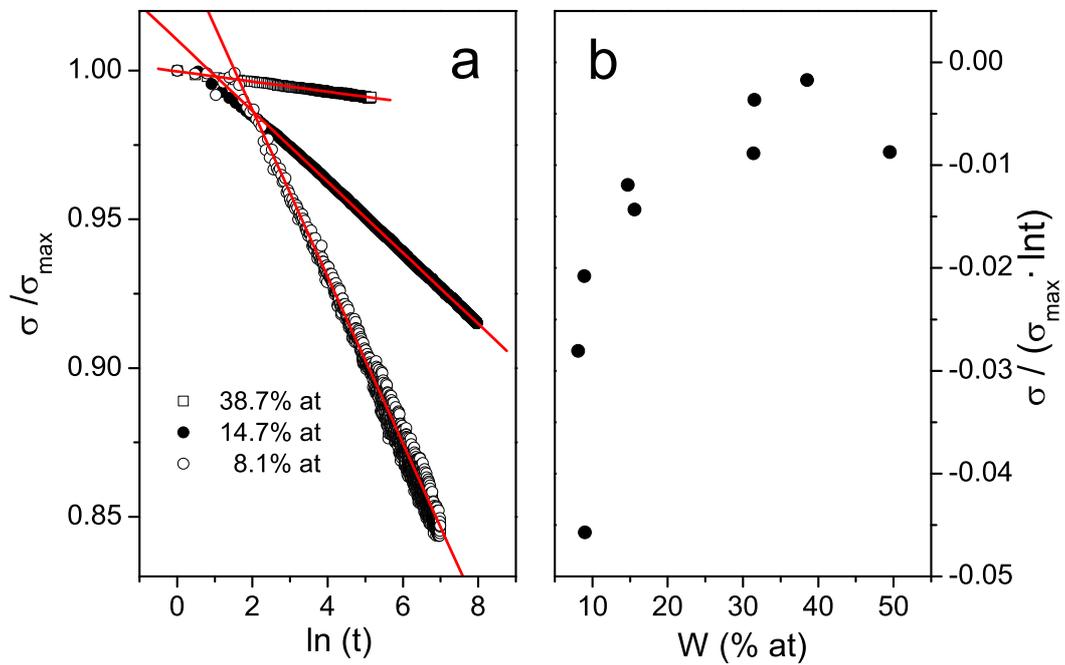}}
\caption{a: Relaxation of the conductivity after the end of the
deposition process. The conductivity decreases with the logarithm
of the time, $\sigma =~b~\cdot~ln(t)$. b: Relaxation rate vs.
metal content. The lower the metal content the faster the
relaxation rate.} \label{relaxation}\end{figure}

The growth process terminates with the simultaneous shut-off of
the electron beam and the precursor gas supply. Correspondingly,
the conductivity starts to decrease (see Fig.~\ref{growth_log}).
The relaxation of the conductivity can be attributed to the
migration of excess electrons injected by the electron beam
towards the electrodes. In Fig.~\ref{relaxation}a we depict the
relaxation for samples $\#$1, $\#$4 and $\#$9. The conductivity
relaxation follows a logarithmic time dependence, i.e.,
$\sigma~=~b~\cdot~ln(t)$. With suitable choice of the parameter
$\textit{b}$, this formula can be used to fit the relaxation  of
all the deposits. The velocity of the relaxation depends on the
deposits' composition. In particular, we notice that deposits with
lower W content relax more rapidly than deposits with higher W
content. This information is summed up in Fig.~\ref{relaxation}b
for all the samples prepared in this work.

\large\textbf{3.3 Measurements during venting the
microscope}\normalsize

\begin{figure}\center{\includegraphics[width=22cm]{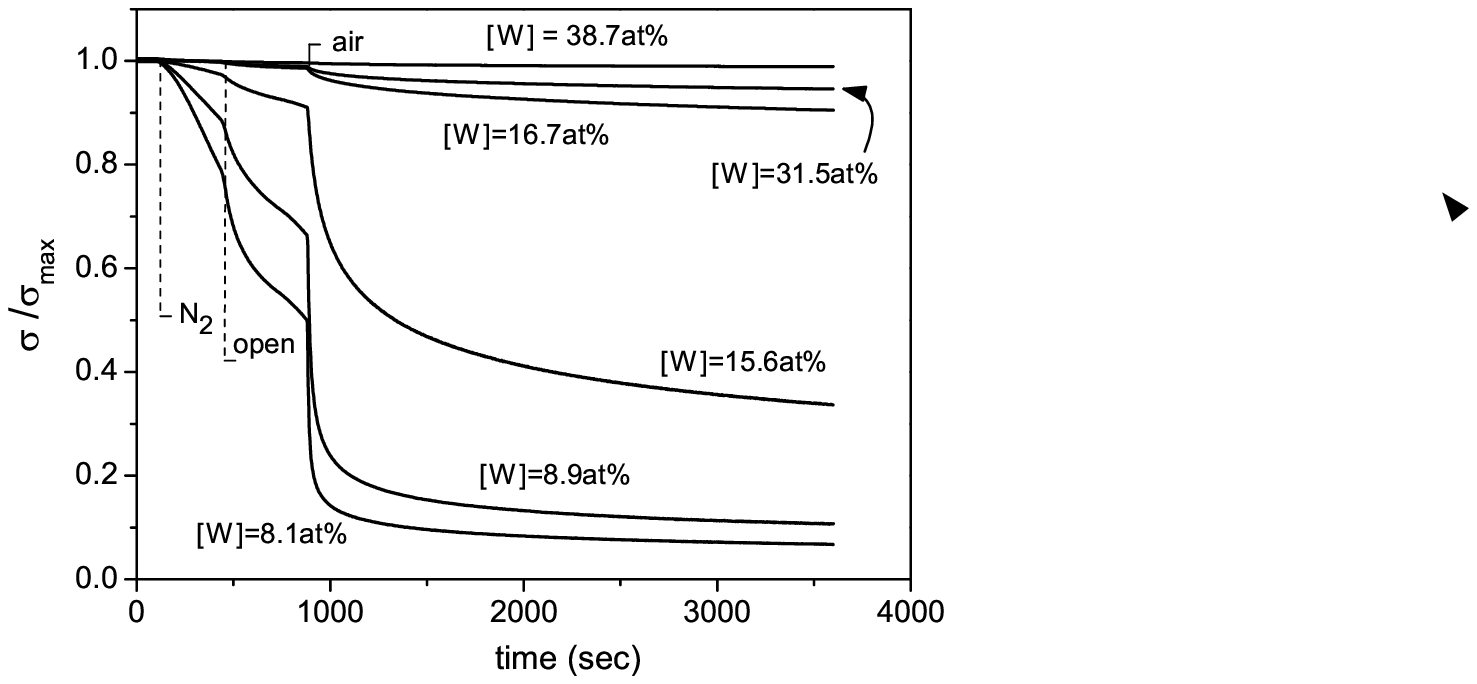}}
\caption{Electrical conductivity vs. time as the SEM is vented.
During the procedure first nitrogen enters the chamber; after
ca.~6~minutes the door of the microscope slightly opens. 8~minutes
later the door is fully open. The conductivity decreases with time
at a rate which depends on the material composition. See text for
details.} \label{exposure}\end{figure}

Transient electrical conductivity measurements during exposure to
air where performed after deposition by venting the electron
microscope. The result of these experiments are summarized in
Fig.~\ref{exposure} where we plot the normalized conductivity
vs.~time for various deposits. The venting procedure starts at
$\textit{t}$=60~s, when $N{_2}$ gas enters the chamber. The
conductivity of the deposits starts to decrease with a rate which
depends on the metal concentration. In particular, the lower is
the metal content the faster is the decrease. The reduction of the
conductivity becomes faster at $\textit{t}$$\approx$400~s, when
the door of the chamber slightly opens and a small flux of air
enters the microscope. At $\textit{t}$$\approx$900~s the door of
the microscope is deliberately opened to fully expose the deposits
to air. At this point the reduction rate of the conductivity
strongly increases. According to Fig.~\ref{exposure}, we divide
the deposits in two groups exhibiting strong and weak degradation
of the conductivity, respectively. To the first group belong the
samples $\#$1, $\#$2, $\#$5 with metal content equal to
8.1~$at\%$, 8.9~$at\%$ and 15.6~$at\%$. After one hour from
opening of the vacuum chamber the conductivity of these deposits
has dropped to 7~$\%$ to 34~$\%$ of its initial value. To the
second group belong samples $\#$6, $\#$7, $\#$9 with 16.7~$at\%$,
31.5~$at\%$ and 38.7~$at\%$. In this case after one hour of
exposure the value of the conductivity lies between 90~$\%$ and
99~$\%$ of the initial value. It is interesting to note that the
degradation rate increases monotonically with decreasing metal
content and, thus, with the beam power used to prepare the
respective samples. Most likely the degradation is due to a
reduction of the tunneling probability within the deposit. In the
discussion section we speculate that the degradation rate can be
linked to the density of the deposit which depends on the beam
power\cite{sawaya}.

\large\textbf{3.4 Post-irradiation conductivity
behavior}\normalsize

\begin{figure}\center{\includegraphics[width=16cm]{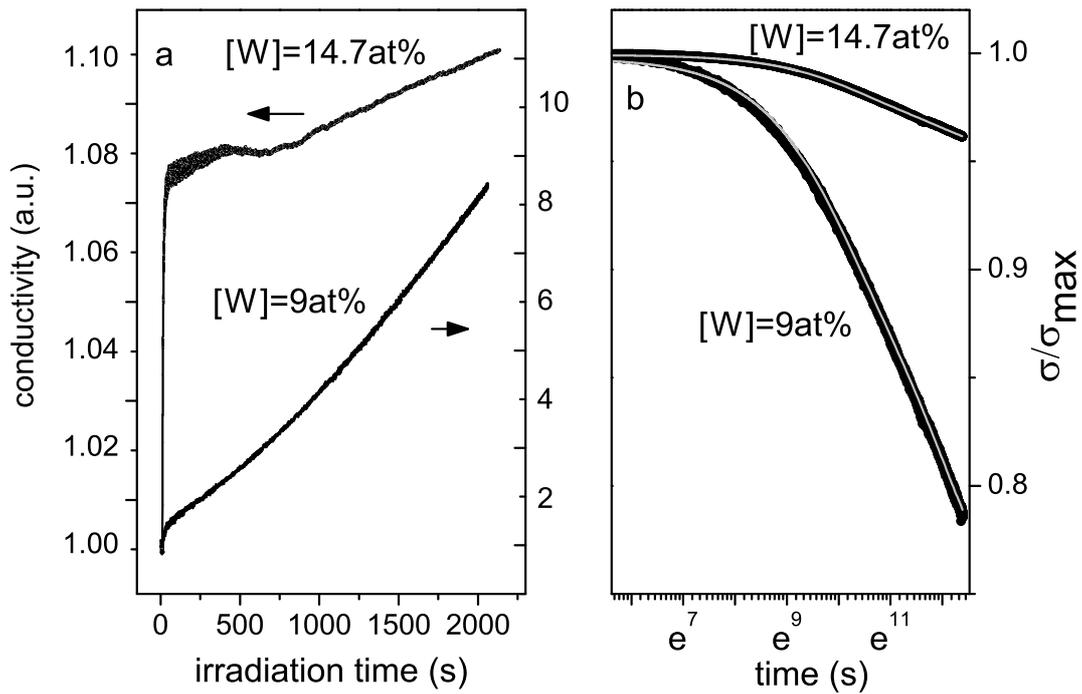}}
\caption{Post-irradiation experiment. Left figure: Conductivity
vs. irradiation time. The increase of the conductivity is the
highest for the sample with lowest W content. Right figure:
evolution of the conductivity after termination of the
irradiation. The decay of the conductivity is described by two
parallel decay channels involving the matrix and the W grains. The
fastest decay is for the deposit with the lowest W content. See
text for details.} \label{irradiation}\end{figure}

In order to study the effect of the beam irradiation on EBID
deposits we consider two samples with W content of 9~$at\%$
(sample~$\#$3) and 14.7~$at\%$ (sample~$\#$4), respectively. After
preparation the samples were irradiated with the same beam power
used during deposition, i.e. 5~nA$\cdot$keV and 10~nA$\cdot$keV,
respectively. In Fig.~\ref{irradiation}a we plot the transient
conductivity vs. the irradiation time. The conductivity of both
samples abruptly increases during the first few seconds of
irradiation. In the following minutes the conductivity increases
monotonically at a much lower rate, which depends on the
composition of the deposit. The conductivity of sample~$\#$3
increases to approximatively eight times its initial value in 2000
seconds of irradiation. In the same time interval the conductivity
of sample~$\#$4 increase only by 10~$\%$ of its initial value. The
increase of conductivity may be attributed to the increase of
charge carriers released during the beam-induced breakage of
carbon-carbon bonds in the amorphous matrix, as was speculated in
Ref.~\cite{bret,babcock}. Sample~$\#$3 shows the largest increase
of conductivity because it is the sample with largest matrix
volume fraction. In Fig.~\ref{irradiation}b we report the
relaxation of the conductivity over a time scale of almost 3 days.
The most remarkable fact is the very low decrease of the
conductivity in comparison with its increase during
post-irradiation. The second remarkable fact is that the
conductivity is not logarithmically dependent on time. Therefore
no relaxation of the conductivity takes place; rather, we suggest
that the conductivity deteriorates because of the exposure to the
rest gases present in the SEM chamber. The phenomena is similar to
the one occurring by exposure to air. However, its time scale is
much longer because the residual pressure ($\approx
4\cdot10^{-6}$mbar) inside the microscope is order of magnitude
lower than in the environment. Note that in the frame of the
"bond-breaking"-model we expect that carbon atoms bond again
leading to a conductivity decrease much faster than the one
observed. Therefore we attribute the increasing conductivity
during irradiation to a permanent chemical transformation of the
deposit.

\large\textbf{3.5 Temperature dependence of the electrical
conductivity }\normalsize

\begin{figure}\center{\includegraphics[width=16cm]{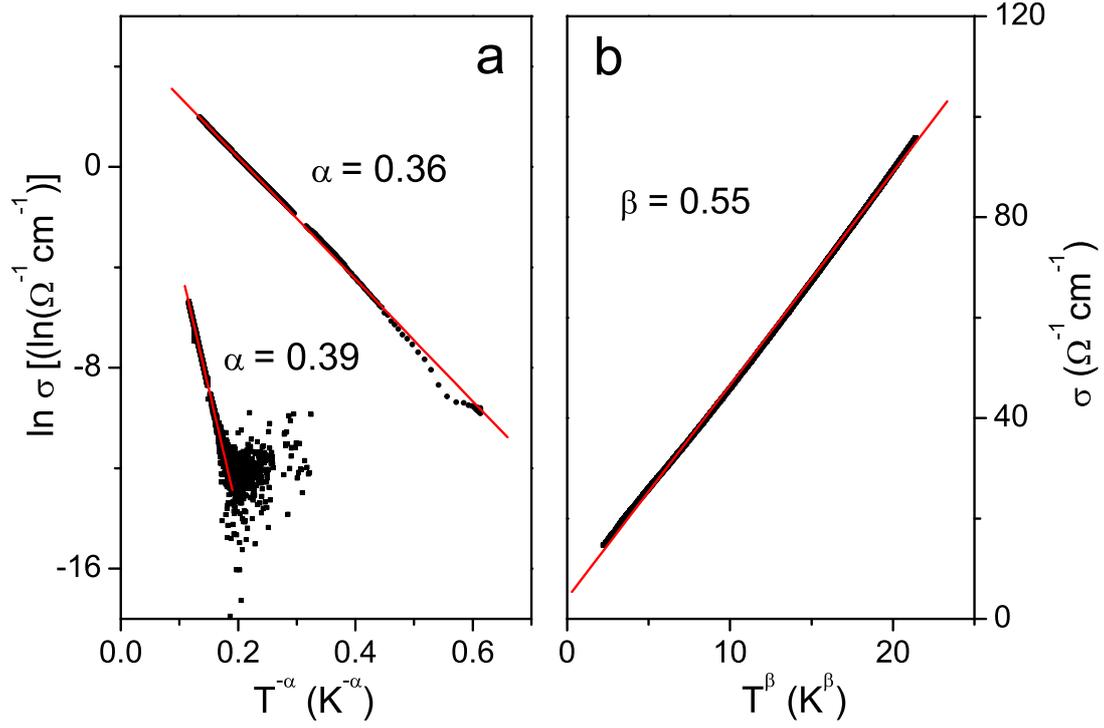}}
\caption{Temperature dependence of the electrical conductivity.
Left figure: the temperature dependence for samples characterized
by variable-range-hopping, ~$\sigma~\propto~exp~(T_{0}/T)^
\alpha$. The exponents $\alpha$=0.39 and $\alpha$=0.36 are found
for samples with W-content of 8.1~$at\%$ (sample $\#$1) and
14.7~$at\%$ (sample $\#$4), respectively. Right figure:
temperature dependence for sample $\#$8 with W-content of
31.8~$at\%$. The conductivity increases with temperature as
$\sigma~=~\sigma{_0}+aT^\beta$, with $\beta=0.55$,
$\sigma{_0}$=5.27 $\Omega^{-1}$cm$^{-1}$ and a=4.08
$\Omega^{-1}$cm$^{-1}$K$^{-1}$.}
\label{conductivity_temperature}\end{figure}

In this section we present the temperature-dependent conductivity
measurements performed for the samples with W content of
8.1~$at\%$ (sample~$\#$1), 14.7~$at\%$ (sample~$\#$4) and
31.8~$at\%$ (sample~$\#$8). For the low conductivity samples~$\#$1
and $\#$4 we choose an applied field between 1143~V/cm and
210~V/cm in order to be able to measure $\sigma(T)$ down to the
lowest temperature accessible in our setup. In order to avoid
irreversible switching in the $\textit{I(V)}$ characteristic
during the measurements we used a current limit. For sample~$\#$8
we choose an applied field of 6.7~V/cm in order to measure
$\sigma(T)$ in the linear regime of the I(V) characteristic. In
Fig.~\ref{conductivity_temperature}a we plot the conductivity vs.
temperature dependence for samples~$\#$1 and $\#$4. Both samples
show the characteristic of systems governed by
variable-range-hopping, i.e. $\sigma~=~\sigma_0~exp~(T_{0}/T)^
\alpha$. In particular, we observed $\alpha=0.28$ and
$\alpha=0.37$ for samples~$\#$1 and $\#$4, respectively. The
behavior in the system with lowest metal content is close to that
one of amorphous carbon films\cite{ambegaokar,hauser}. In that
case the tunneling takes place between localized sites in the
limit of negligible Coulomb blockade effects
($\alpha=0.25$)\cite{ambegaokar}. The value of $\alpha$ obtained
for sample~$\#$4 may suggest both, a contribution to the tunneling
between the trap sites in the matrix and between the metal grains;
the latter mechanism is described by the exponent
$\alpha=0.5$\cite{abeles,sheng}. In
Fig.~\ref{conductivity_temperature}b we plot the temperature
dependence of the electrical conductivity for sample~$\#$8. We
find that the conductivity follows a power law over the complete
temperature range of our measurement (1.8~K$\leq T \leq$265~K)
with~$\sigma~=\sigma_0+a~T^\beta$, with $\beta=0.55$. Such a
behavior is similar to the one recently reported in
Ref.~\cite{huth} and interpreted by means of a tunneling
percolation model in the limit of large inter-grain
tunneling~\cite{huth,beloborodov}.

\begin{center}
\large\textbf{4. Discussion}\normalsize
\end{center}

\begin{figure}\center{\includegraphics[width=8cm]{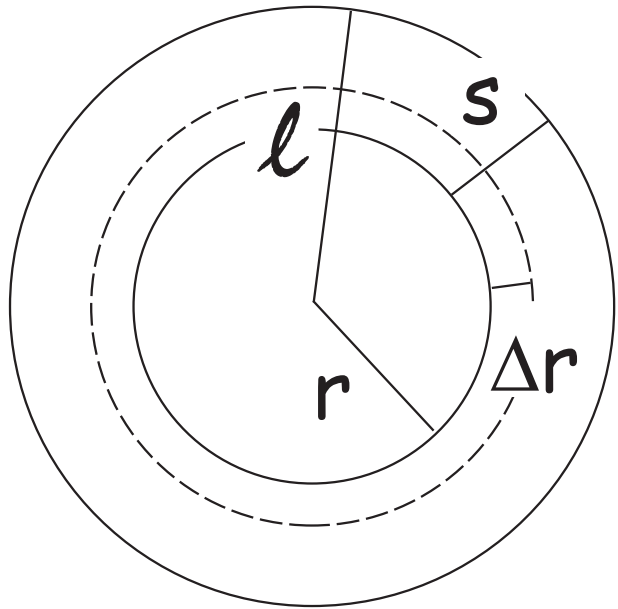}}
\caption{"Shell-model" used to model the time-dependent
conductivity during the growth of a deposit. The model assumes
that the metal W-particles have initially a radius $\textit{r}$.
The radius growth by $\Delta r$ in the time $\Delta t$ during the
deposition, because of e-beam stimulated diffusion of W atoms.
$\textit{s}$ is the separation between neighboring particles,
$\textit{l}$ is the sum of $\textit{s}$ and $r$.}
\label{shell_model}\end{figure}

\begin{figure}\center{\includegraphics[width=16cm]{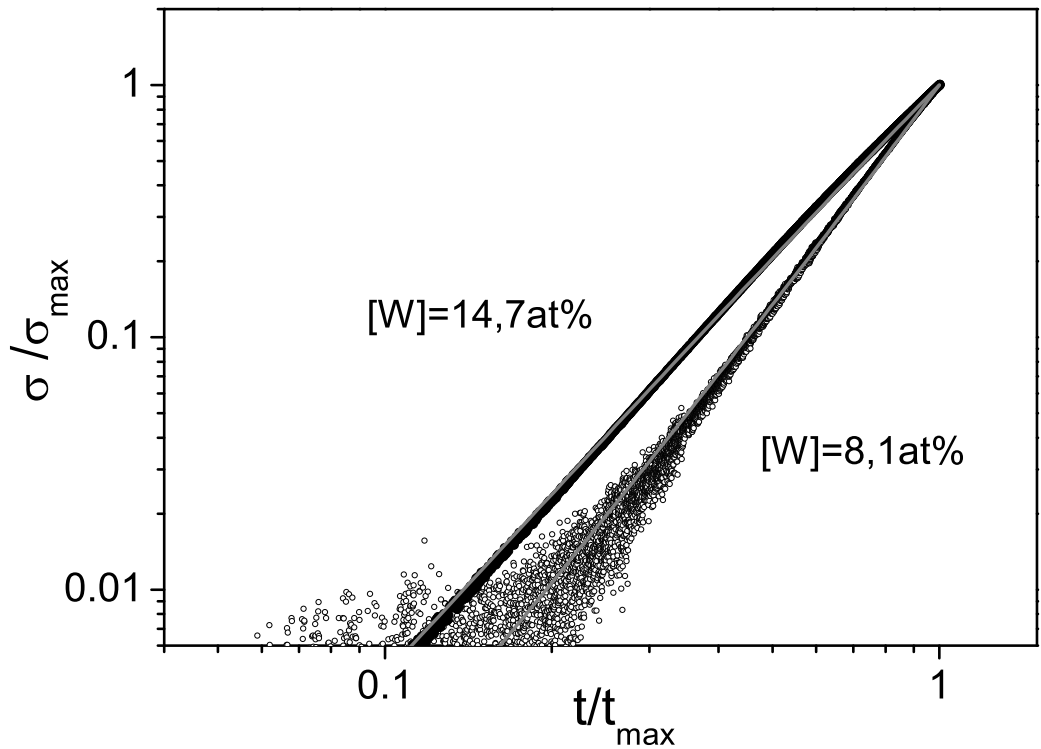}}
\caption{Conductivity vs. time during the growth process. The fit
to the experimental data are made in the frame of the
variable-range-hopping theory, see eq.(~$\ref{eqhop}$). The time
dependence of the conductivity is implicit in the localization
length $\xi$ and in the effective activation energy $\textit{W}$,
according to the shell model of Fig.~$\ref{shell_model}$.}
\label{sigma_diameter_time}\end{figure}

A linear increase of the time-dependent conductivity can be
explained by supposing a linear time dependent growth of the
thickness. However, the measurements performed during growth show
an increase of the conductivity vs. time faster than linear.
Therefore, an additional mechanism which contributes to the
enhancement of the conductivity has to be considered. Possible
additional mechanisms for the increase of the conductivity
growth-rate include an enhancement of the electron carrier density
and of the metal-particles' size during deposition, both due to
the electron beam irradiation. In the following we relate the time
dependence of the conductivity to the increase of the metal
particles' size by means of a "shell-model". We consider a sphere
of radius $\textit{r}$ and volume $v_s$ representing a
metal-particle, see Fig.~\ref{shell_model}. Coordination sphere of
radius $\textit{l}$ and volume $v_l$ surrounds the first one. The
radius difference $\textit{s=l-r}$ represents the separation
between two neighboring particles. The ratio $v=v_s/v_l$ gives the
metal-volume fraction of the system. It follows that the diameter
of the metal particle is $d=2~r=2~l~v^\frac{1}{3}$. We suppose
that at time $\textit{t=t$_0$}$ the deposit is constituted by
particles with the same radius $\textit{r}$. The particles of this
initial deposit increase their volume because of electron beam
stimulated diffusion of W-atoms. In general, the growth rate of
the particle size is described by the Lifshitz-Slyozov coarsening
law $r\sim t^\frac{1}{3}$\cite{LS}. However, also lower growth
rates with $r\sim t^\frac{1}{n}$ with n$>$3, have been
reported\cite{goldfarb,muratov}, which indicate diffusion limited
mass transport. Known the diameter $\textit{d}$, the separation
$\textit{s}$ and the growth rate, the time-dependence conductivity
in the VRH regime can be deduced from equation ($\ref{eqhop}$)

\begin{equation}
  \sigma\sim e^ {-2\frac{s}{\xi}-\frac{W}{k_BT}}
 \label{eqhop}
 \end{equation}

where $\xi$ is the localization length, $\textit{k$_B$}$ the
Boltzmann constant, $\textit{T}$ the temperature and $\textit{W}$
the effective activation energy for charge
transport\cite{adkins,entin}. The time dependence of the
conductivity is implicit in the localization length and in the
activation energy, both dependent on the particle diameter. In
particular, $\xi=\xi_{mx}(1+d/s)$, with $\xi_{mx}$ being the
localization length of the matrix\cite{adkins}. The localization
length is obtained from the relation
$\xi_{mx}=2.8e^2/4\pi\epsilon\epsilon_0k_BT_0(1+d/s)$\cite{tran},
where $\epsilon$ is the dielectric constant of the deposit and
$\textit{T$_0$}$ a constant deducible from the
temperature-dependent conductivity measurements. From the
experiments one also extracts the activation energy
$\textit{W}$=$\alpha~kT(T_0/T)^\alpha$\cite{adkins}. From
Fig~\ref{conductivity_temperature}a it results $T_{0}= 1.28\cdot
10^5$~K and $T_{0}= 6.4\cdot 10^3$~K, for samples $\#1$ and $\#4$,
respectively. The values of the dielectric constant necessary to
calculate $\xi_{mx}$ are chosen considering that for a
carbon-matrix $\epsilon\approx$4\cite{grill} and that with
increasing metal volume fraction, the dielectric constant of a
metal-insulator composite increases\cite{efros}. By using for
$\epsilon$ the values 8 and 12, it follows $\xi_{mx}=0.02$~nm and
$\xi_{mx}=0.1$~nm, $\xi$=0.03~nm and $\xi$=0.61~nm for samples
$\#1$ and $\#4$, respectively. For a comparison note that the
value of the localization length of amorphous carbon thin films
lies between 0.1~nm and 1.2~nm\cite{frauenheim}. In
Fig.~\ref{sigma_diameter_time} we report the time-dependent
conductivity measurements for samples $\#1$ and $\#4$. In order to
fit the experimental data, we choose a linear increase of the
metal volume fraction $\textit{v}$ from 0.04 to 0.06 and from 0.04
to 0.35 for samples $\#1$ and $\#4$, respectively. The best fit
was obtained with $\textit{n}$=9.8 and $\textit{n}$=6.3,
0.61$\leq$~s/l~$\leq$0.65 and 0.28$\leq$~s/l~$\leq$0.66,
respectively for samples $\#1$ and $\#4$. These choices for
$\textit{l}$ and $\textit{v}$ lead to
0.55~nm~$\leq$~s~$\leq~$0.59~nm and
0.42~nm~$\leq$~s~$\leq$~0.99~nm, 0.62~nm~$\leq$~d~$\leq$0.71~nm
and 1.03~nm~$\leq$~d~$\leq$~2.16~nm, respectively for the two
samples. In conclusion, the experimental data on the growth
process can be well described within our "shell-model" by using a
coarsening growth rate which depends on the deposit's metal
content and, thus, on the beam parameters. We find that in the
deposit with lowest metal content (sample $\#1$, n=9.8) the size
of the metal particle increases in time with the lowest rate, as
one may have expected. The coarsening law of Lifschitz-Slyozov is
not reached by the range of beam parameters used for sample $\#1$
and $\#4$. We believe that for higher deposition rates, such
regime may be reached. It has to be noted that we cannot describe
the behavior of sample $\#9$ by means of the previous modelling
because the parameters $T_0$ and $\xi$ cannot be deduced from the
temperature dependence of the conductivity of
Fig.~\ref{conductivity_temperature}b. However, it is quite
possible that in the first stages of the deposition process the
electron transport may be described by the VRH mechanism. In this
case, the growth rate of the particle size of sample $\#9$ shall
be much higher than the one of samples $\#1$ and $\#4$, being
close to the Lifschitz-Slyozov theory. Indeed, an hint that this
may be the case is given by the different power law, which is
$\sigma$=~$\textit{t}$~${^4}$ for sample $\#9$ in the first
seconds of deposition, while it is
$\sigma$=~$\textit{t}$~${^{2.86}}$ for sample $\#1$, during the
whole range of deposition time. It is important to note that the
"shell-model" assumes the presence of nanocrystal particles in the
deposit. According to the TEM measurements of Hoyle $\textit{et
al.}$\cite{hoyle} this assumption is fully justified for deposits
prepared with high doses. For deposits prepared with a dose per
scan smaller than 500 C/m$^2$ the authors of Ref.~\cite{hoyle}
found amorphous material. However, the presence of very small
crystals not detectable from TEM measurements cannot be excluded.
Indeed, the sharp fall of the conductivity with the decrease of
the doses (see inset in Fig.~\ref{composition}) may also be
explained by the Coulomb blockade of small nanocrystals at room
temperature.

In granular materials it is observed that in the low conductivity
tunneling regime the temperature dependence of the electrical
conductivity is of the form ~$\sigma~\sim~exp~(T_{0}/T)^ \alpha$,
with 0.25$\leq\alpha\leq$0.5. For our samples we find
0.36$\leq\alpha\leq$0.5 with metal content $\leq19\%$, which may
indicate a tunneling mechanism either involving localized sites in
the matrix or directly between the metal grains. However, in order
to have tunneling the typical hopping distance for the electron,
$r^*=(\frac{e^2\xi}{4\pi\epsilon\epsilon{_0}k{_B}T})^{0.5}$, has
to be larger than the distance $\textit{s}$ between the localized
states\cite{tran}. For the samples $\#1$ and $\#4$ we obtain
$r^*$=0.46~nm and $r^*$=1.69~nm, respectively. By comparing these
numbers with the ones for $\textit{s}$ given above by the
"shell-model" we find that in our deposits direct tunneling
between grains appears to be possible. On the other hand,
tunneling between localized sites in the matrix is not favorable.
However, this does not exclude conduction inside the matrix which
may also take place between localized matrix sites bridged by
metal grains. As a consequence in the following we speculate that
the transport mechanism in our deposits takes place by means of
two channels: directly between grains and indirectly between
localized sites in the matrix through a grain. In order to
describe the transport mechanism we use the following formula:

\begin{equation}
  \sigma=a\cdot \sigma_{01}\cdot
e^{-(\frac{T_{01}}{T})^{\frac{1}{4}}}+b\cdot \sigma_{02}\cdot
e^{-(\frac{T_{02}}{T})^{\frac{1}{2}}} + \sigma_{03}\cdot
e^{-(\frac{T_{03}}{T})^{\delta}}
 \label{eq1}
 \end{equation}

The first two terms of equation ($\ref{eq1}$) take into account
the tunneling of electrons between trap sites in the matrix and
between grains, respectively. The factor $\textit{a}$ gives the
volume concentration of the matrix, while $\textit{b}$ refers to
the concentration of the metal. The third term of the equation
takes into account the hopping of electrons between trap sites and
grains. The temperature $T_{03}$ and factor $\delta$ are free
parameters. By fitting the experimental curves with
equation~($\ref{eq1}$), we obtain the volume concentration of the
matrix and of the metal. We made the fit to the two curves of
Fig.~\ref{conductivity_temperature} ("`fit~1"' and "`fit~2"'), for
which we have W content of 8.1at$\%$ and 14.7at$\%$, associated to
$\alpha=0.39$ and $\alpha=0.36$, respectively. For the fit we set
$\sigma_{01} = 10$~$\Omega~^{-1}$~cm$^{-1}$, $\textit{T$_{01}$}$ =
10$^7$~K \cite{frauenheim}, $\sigma_{02} =
27$~$\Omega~^{-1}$~cm$^{-1}$ and $\textit{T$_{02}$}$ =
2070~K\cite{huth}. For the third term of eq.~($\ref{eq1}$) we
choose $\sigma_{03} = 15$~$\Omega~^{-1}$~cm$^{-1}$,
$\textit{T$_{03}$}$ = 1.3$\cdot$10$^4$~K (fit~1) and
$\textit{T$_{03}$}$ = 7$\cdot$10$^3$~K (fit~2). The best fit was
obtained for $a=0.06$, $b=0.94$ and $\delta= 0.38$ (fit~1) and
$a=0.35$, $b=0.65$ and $\delta= 0.38$ (fit~2). We can compare the
fractional volume obtained by the fit with the one deduced by EDX
measurements. The fractional volume occupied by the metal
particles depends on the density of the metal and of the matrix,
which is not well known. The lower and upper limit corresponding
to a density of 1.8~g/cm$^3$ (graphite-like amorphous) and of
3.5~g/cm$^3$ (diamond-like) lead for the W-content of 8.1at$\%$ to
a fractional volume between 12vol$\%$ and 20vol$\%$\cite{huth}.
From our fit we obtain the volume of 6$\%$, which would indicate a
lower density than a graphite-like matrix. This difference may be
due to the presence of oxygen atoms inside the carbonaceous
matrix. By means of in situ mass measurements the density of the
matrix of a W-based deposits from W(CO)$_{6}$ has been related to
the electron beam parameters used during deposition\cite{sawaya}.
In general, the density of the matrix increases with the power of
the electron beam. The authors measured a matrix density between
0.29~g/cm$^3$ and 0.88~g/cm$^3$ corresponding to a W-volume
concentration between 11.1~vol$\%$ and
6.4~vol$\%$\cite{sawaya,nishio}. These results where obtained for
beam voltages between 5~keV and 15~kev, but with a beam current
much higher than the one used in our depositions. For a system
with W-content of 14.7at$\%$, we calculate a fractional volume
between 19vol$\%$ and 33vol$\%$ for graphite- and diamond-like
matrix, respectively. The value of 35~vol$\%$ obtained within our
modelling is close to that for a diamond-like matrix. Of course
the matrix of our deposit is non-crystalline and does also contain
oxygen. In conclusion, we find clear evidence of an increase of
the density of the matrix by increasing the electron beam power.
This result is in accord with the one obtained for amorphous
carbon deposits obtained by electron-beam-induced
deposition\cite{sawaya,nishio}.

After terminating the growth process the conductivity starts to
decrease with the logarithm of the time, see
Fig.~\ref{relaxation}. The relaxation can be attributed to the
migration of the charge carriers out of the sample. In general, in
order to describe the relaxation of a system toward equilibrium a
stretched exponential of the form $\sigma\sim
exp~(-t/\tau)^{\beta}$, with 0$\leq \beta \leq$1 can be
used\cite{kohlrausch}. In the case of $\beta$=1 the the Debye
relaxation is obtained, which is valid for systems with weak
dynamic correlations. In the case of hopping conductivity, this
corresponds to a system with no Coulomb interactions. If the
system presents strong dynamic interactions, the law of
Kohlrausch\cite{kohlrausch} can be obtained by considering a
serial relaxation of hierarchies of correlated
systems\cite{palmer}. In case of hopping conductivity, this
corresponds to a system with strong Coulomb interactions,
dynamically described by the tunneling of an electron into a
localized state, which is possible if a second electron tunnels
into another localized state (co-tunneling). The logarithmic decay
of the variable describing a system finds an explanation in the
frame of hierarchically constrained dynamics\cite{brey}. In
particular, in the theoretical study of Ref.~\cite{tsigankov} it
is mentioned that the conductivity of systems governed by hopping
conduction can exhibit a logarithmic time dependence, as we find
experimentally. Finally, we observed an increase of the relaxation
rate at lower W-concentration. We attribute this behavior to the
increased number of localized sites in the matrix and, therefore,
to an higher probability for the electron to tunnel into a free
localized state.

\begin{figure}\center{\includegraphics[width=16cm]{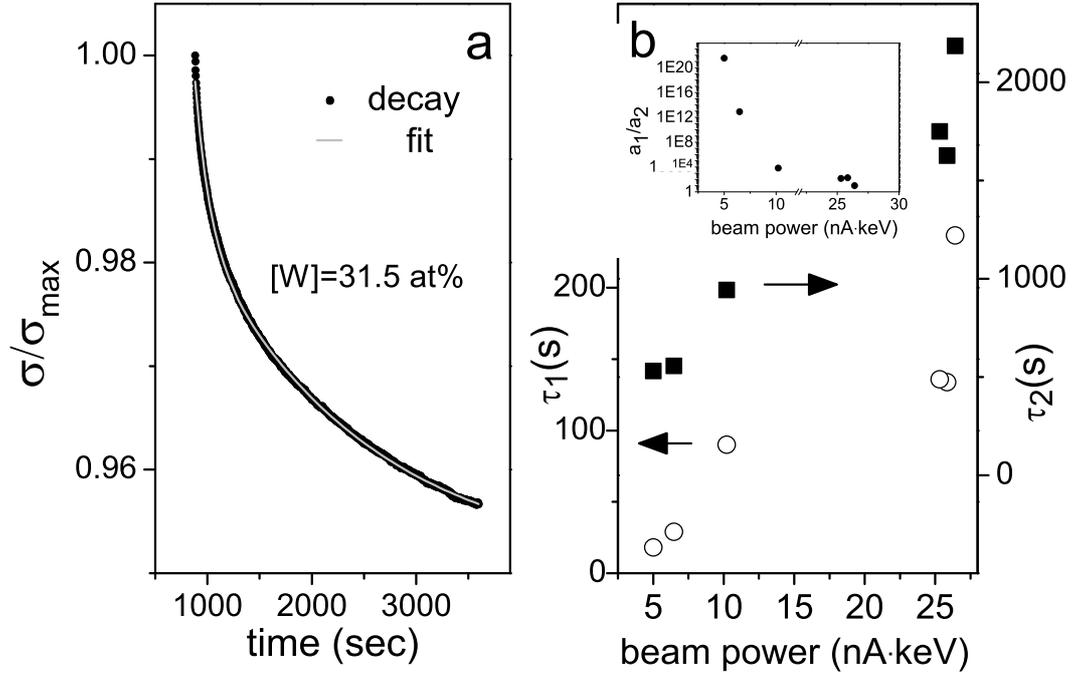}}
\caption{Decay of the conductivity during exposure to air. Left
figure: the decay, as exemplary shown for sample $\#$7, is
described by the sum of two exponential decay channels according
to $\sigma$~=~a$_1$~exp(t/$\tau_1$)+a$_2$~exp(t/$\tau_2$). Each
decay channel is associated to a tunneling mechanism: either
between grains and trap sites in the matrix (decay constant
$\tau_1$) or between metal grains (decay constant $\tau_2$). Right
figure: value of the decay constants vs. beam power. The inset
shows the ratio between the weight of the two tunneling
mechanisms: a$_1$/a$_2$. For deposits prepared at small beam power
the tunneling involving the matrix is dominant.}
\label{decay_exposure}\end{figure}

The increase of the density of the matrix with the electron beam
current (see above) can be used to interpret the conductivity
measurements obtained during air exposure. The degradation rate
$\sigma'$ decreases as one considers deposits prepared at larger
beam current (power) and, thus, with the density of the matrix. In
order to better describe the degradation process, we fit the
experimental data of Fig.~\ref{exposure}, from the time where the
door of the microscope is opened ($\textit{t}$$\approx$900~s),
with a sum of two exponential functions: $\sigma\propto$
a$_1$~exp(t/$\tau_1$)+a$_2$~exp(t/$\tau_2$). In
Fig.~\ref{decay_exposure}a we plot a typical decay fitted with
such a curve. In Fig.~\ref{decay_exposure}b we report the decay
constants $\tau_1$ and $\tau_2$ for the deposit measurements
depicted in Fig.~\ref{exposure} as a function of the beam power
used to prepare the samples. Both decay constants increase with
the beam power with ($15 \leq \tau_1 \leq 236$)~s$^{-1}$ and ($510
\leq \tau_2 \leq 2185$)~s$^{-1}$. As one observes from the inset
of Fig.~\ref{decay_exposure}b the weight of the first term of the
exponential sum is larger than the weight of the second one
(a$_1$$>$a$_2$). In particular the ratio strongly increases for
small values of the beam power, which corresponds to deposits
mainly constituted by the matrix. Therefore we associate the first
term of the exponential sum to the matrix. The decay constant
$\tau_1$ is associated to the decrease of the tunneling
probability within the matrix, or between matrix and grains in
view of our previous discussion. The increase of the decay
constant $\tau_1$ with beam power is interpreted as due to the
increase of the matrix density. The second term of the exponential
sum may describe the decay of the tunneling probability between
metal grains. The increase of the decay constant $\tau_2$ with
beam power may be interpreted as due to the increase of the grain
size.

Similar considerations are useful in order to analyze the behavior
of the deposits after irradiation. Also in this case the
conductivity can be described by the sum of two exponential
functions. The two decays are independent processes linked to the
decrease of the tunneling probability within the matrix and
between the metal grains. The decay constants of sample $\#$4 and
sample $\#$3 are $\tau_1$=1.91$\cdot 10^4$~s, $\tau_1$=2.33$\cdot
10^4$~s and $\tau_2$=1.66$\cdot 10^5$~s, $\tau_2$=1.6$\cdot
10^5$~s, respectively. These values are 2 to 3 orders of magnitude
larger than those obtained by venting the system. From the
literature it is known that the presence of residual oxygen and
water molecules inside a SEM is sufficient to oxidize W-based
granular metals prepared from the tungsten hexafluoride (WF$_6$)
precursor\cite{klein}. Similarly, the oxidation of the deposits is
most likely the reason for the observed decrease of the
conductivity that we measured inside the SEM. As can be expected,
the degradation takes place on a much longer time scale in
comparison to the exposure to air.

\begin{center}
\large\textbf{5. Conclusions}\normalsize
\end{center}

Probably, the most relevant technological aspect concerning our
study is the investigation of the behavior of W-based granular
metals during exposure to air. We have found that the degradation
of the system strongly depends on the composition. In particular,
after one hour of exposure to air the remaining conductivity
varies between 7$\%$ and 99$\%$ of its initial value for deposits
with W-content in the range 8.1at$\%$$\leq$W$\leq$38.7at$\%$.
Transient electrical conductivity measurements indicate that this
dependence is due to the different densities of the deposits.
Furthermore, the decay of the conductivity involves the matrix and
the W-metal grains on two different time scales. In conclusion, in
view of the possible employment of EBID W-based deposits for
applications the composition of the composite has to be carefully
chosen to minimize ageing effect. In this direction it is
remarkable that after one year from deposition two samples with
W-content of 34$\%$ and 19$\%$, respectively, show a decrease of
the conductivity only of a factor between two and four. However,
covering the deposit with a protective SiO$_x$, layer like
tetraethylorthosilicate (TEOS) precursor, may be necessary to
prevent long time scale degradation of the electrical
properties\cite{botman}.

By means of transient electrical conductivity measurements we have
monitored the growth process of W-based deposits. In order to
explain the non-linear growth of the conductivity vs. time we have
used a "shell-model" based on the hypothesis that the diameter of
the W-metal particles grows during deposition. Within the
modelling we have estimated the average particle size in the final
deposit, which ranges between 0.71~nm and 2.16~nm for deposits
with W content of 8.1$at\%$ and 14.7$at\%$. In would be
interesting to integrate this analysis with transmission electron
microscopy investigations.

Furthermore, we have investigated the electrical transport
mechanism as a function of the deposits' composition by means of
conductivity vs. temperature measurements. From the analysis of
the experimental data we have speculated that the electrical
transport takes place by means of two parallel channels involving
both the direct tunneling between metal-particles and the indirect
tunneling between localized sites in the matrix bridged by metal
particles. As a consequence of this investigation information
about the matrix density can be deduced, which is useful to
interpret the data of the conductivity degradation under exposure
to air. Finally, we have observed that the relaxation of the
charge carriers after termination of the deposition process
follows a logarithmic time dependence. Such a behavior suggest
that at the end of the deposition the electron transport in the
deposits is dominated by strong Coulomb interaction and that the
relaxation can be described in the framework of hierarchically
constrained dynamics commonly used to describe strongly
interacting glassy materials.

In conclusion, we have shown that by means of $\textit{in situ}$
transient electrical conductivity measurements very  useful
information about the electrical transport mechanism, the
microstructure and the chemical and physical stability of W-based
granular materials can be obtained. It shall be interesting to
extend this measurements to materials grown with other precursor
gases and to further these studies with TEM investigations.

\begin{center}
\large\textbf{Acknowledgments}\normalsize

Financial support by the \textit{NanoNetzwerkHessen (NNH)} and by
the \textit{Bundesministerium f\"ur Bildung und Forschung (BMBF)}
under grant 0312031C is gratefully acknowledged.

\end{center}

\newpage

\end{document}